\documentclass[letterpaper,showpacs,preprintnumbers,twocolumn,eqsecnum,superscriptaddress,aps,nofootinbib]{revtex4}
\usepackage{amssymb}
\usepackage[centertags]{amsmath}
\usepackage{txfonts}
\usepackage{epsfig}
\usepackage{bm}
\usepackage{color}
\usepackage{graphicx,graphics}
\usepackage{subfigure}
\usepackage{multirow}
\usepackage{float}
\usepackage[pdfstartview=FitH]{hyperref}
\hypersetup{colorlinks=true, citecolor=blue, linkcolor=blue,filecolor=black,urlcolor=blue}
\allowdisplaybreaks[4]
\usepackage{slashed}
\usepackage{ulem}
\usepackage{cancel}
\usepackage{booktabs}
\usepackage{multirow}
\usepackage{pdflscape}

\usepackage{hhtensor}
\usepackage{rotating}

\begin{document}
\title{Azimuthal and spin asymmetries in $e^+e^-\to V \pi X$ at high energies and 3D fragmentation functions}

\author{Kai-bao Chen}
\affiliation{School of Physics \& Key Laboratory of Particle Physics and Particle Irradiation (MOE), Shandong University, Jinan, Shandong 250100, China}

\begin{abstract}
We present the systematical decomposition results of three-dimensional (3D) fragmentation functions (FFs) from parton correlators for spin-1 hadron.
We choose one of the best process $e^+e^-\to V\pi X$
to study the 3D and tensor polarization dependent FFs.  
By making a general kinematic analysis we show that the cross section is expressed by 81 independent structure functions,
and get the results of the azimuthal and spin asymmetries.
We also present the parton model results for this process
and express the results in terms of the 3D FFs.
Based on this formalism, we also give the numerical results of hadron longitudinal polarizations and predict the energy dependence.
\end{abstract}


\maketitle

\section{Introduction} \label{sec:introduction}
Parton distribution functions (PDFs) and FFs are both important inputs for describing high energy reactions.
PDFs are used to describe hadron structure and the FFs describe the hadronization mechanism. 
They should be studied in parallel to promote our understanding of non-perturbative quantum chromodynamics (QCD).
In inclusive high energy reactions, only longitudinal momentum dependent PDFs and/or FFs are involved. 
For describing semi-inclusive reactions,
3D or transverse momentum dependent (TMD) PDFs and/or FFs are needed
to explain the phenomenon such as azimuthal asymmetries of the hadron.
This will generalize our knowledge of hadron structure and hadronization mechanism to three dimensional case.
See e.g. Ref.~\cite{Chen:2015tca} for a recent brief review.

Based on the parallel talk given at the 22nd International Spin Symposium and Ref.~\cite{Chen:2016moq},
in this proceeding we first summarize the results for 3D FFs defined via quark-quark correlator for spin-1 hadrons systematically.
Then we choose the process $e^+e^-\to V \pi X$ which is one of the best places to study these 3D and tensor polarization dependent FFs. 
We give a general kinematical analysis of this process and get the differential cross section expression in terms of structure functions.
We also present the parton model calculation results for this process in leading order perturbative QCD and up to twist-3.
We get the structure functions results as well as the hadron azimuthal asymmetries and polarizations in terms of the convolution of the TMD FFs.
We also present the numerical results of energy dependence of the hadron polarizations based on this formalism.

\section{TMD FFs from quark-quark correlator} \label{sec:FFs}
A complete parameterization for the TMD PDFs for spin-$1/2$ hadron can be found e.g., in \cite{Goeke:2005hb}.
Here, we summarize the results for TMD FFs of spin-$1$ hadron production case.

The leading contribution of the TMD quark fragmentation is defined by the TMD quark-quark correlator, i.e.,
\begin{align}
\hat\Xi&^{(0)}(z,k_{\perp};p,S) =
\sum_X \int \frac{p^+d\xi^-}{2\pi} d^2{\xi}_\perp e^{-i(p^+\xi^-/z - \vec{k}_{\perp} \cdot \vec{\xi}_\perp)}  \nonumber\\
&\times  \langle 0| \mathcal{L}^\dag (0;\infty) \psi(0) |p,S;X\rangle \langle p,S;X|\bar\psi(\xi) \mathcal{L}(\xi;\infty) |0\rangle, \label{Xi0}
\end{align}
where $k$ and $p$ denote the 4-momenta of the quark and the hadron respectively, $S$ denotes the spin of the hadron; 
$z=p^+/k^+$ is the light-cone momentum fraction of the hadron;
$\mathcal{L}(\xi;\infty)$ is the famous gauge link making this definition gauge invariant.
The correlator satisfies hermiticity and parity conservation,
which are important constraints when we decompose it.
TMD FFs are obtained from Eq.~(\ref{Xi0}) by decomposing it in two steps.
First, we expand $\hat\Xi^{(0)}(z,k_{\perp};p,S)$ in terms of the $\Gamma$-matrices, i.e., (we omit the arguments for simplicity)
\begin{align}
\hat\Xi^{(0)} =  \Xi^{(0)} + i\gamma_5 \tilde\Xi^{(0)} + \gamma^\alpha \Xi_\alpha^{(0)} + \gamma_5\gamma^\alpha \tilde\Xi_\alpha^{(0)} + i\sigma^{\alpha\beta}\gamma_5 \Xi_{\alpha\beta}^{(0)}. \label{XiExpansion}
\end{align}
The corresponding coefficient functions are Lorentz scalar, pseudo-scalar, vector, axial-vector and tensor respectively.
Second, we make a Lorentz structure decomposition of these coefficient functions using the available variables at hand. 
The coefficients are expressed as the sum of the basic Lorentz covariants multiplied by scalar functions of $z$ and $k_{F\perp}^2$,
which are defined as TMD FFs.

Take the unpolarized part as an example. 
The only Lorentz scalar we can build is $p^2$ or the hadron mass $M$; no pseudoscalar can be built;
we have three vectors $p_\alpha$, $k_{\perp\alpha}$, $n_\alpha$ 
and one axial-vector $\tilde k_{\perp\alpha} \equiv \varepsilon_{\perp \beta\alpha} k_{\perp}^\beta$;
three tensors $p_{[\rho}\tilde k_{\perp\alpha]}$, $\varepsilon_{\perp\rho\alpha}$ and $n_{[\rho}\tilde k_{\perp\alpha]}$ are available.
So we can immediately write down the decomposition of the TMD FFs as
\begin{align}
& z\Xi^{U(0)} = ME,\quad z\tilde\Xi^{U(0)} =0, \quad z\tilde\Xi_\alpha^{U(0)} = -\tilde k_{\perp\alpha} G^\perp,\\
& z\Xi_\alpha^{U(0)} = p^+ \bar n_\alpha D_1+ k_{\perp\alpha} D^\perp + \frac{M^2}{p^+}n_\alpha D_3,\\
& z\Xi_{\rho\alpha}^{U(0)} = -\frac{p^+}{M} \bar n_{[\rho}\tilde k_{\perp\alpha]} H_1^\perp + M\varepsilon_{\perp\rho\alpha} H - \frac{M}{p^+} n_{[\rho}\tilde k_{\perp\alpha]} H_3^\perp.  \label{eq:XiUT}
\end{align}
Here and in the following, we will omit the arguments $(z,k_{\perp})$ for the FFs for simplicity.
We see that there are two leading twist TMD FFs, i.e., the number density $D_1$ and Collins function $H_1^\perp$,
and two corresponding twist-4 addenda $D_3$ and $H_3^\perp$. Other four are twist-3 FFs.
To save space, for the vector polarization and tensor polarization dependent parts, 
we refer interested readers to Ref.~\cite{Chen:2016moq} for the complete decomposition results 
which include 72 TMD FFs, as well as the discussions of the naming system and the properties of the TMD FFs.

Higher twist FFs can also be defined via quark-gluon(s)-quark correlators.
However, they are often not independent because of QCD equation of motion $\gamma \cdot D(y)\psi(y)=0$.
We also refer to Ref.~\cite{Chen:2016moq} for the complete decomposition results of the twist-3 TMD FFs from quark-gluon-quark correlator.
Here, we just emphasize that a unified relationships 
between the twist-3 FFs defined from quark-quark and quark-gluon-quark correlator can be obtained as,
$z(D^K_{dS} +G^K_{dS}) = D^K_S+iG^K_S$ for the chiral-even part 
and $2z(H^K_{dS} + \frac{k_{\perp}^2}{2M^2} H^{K'}_{dS}) = H^K_S + \frac{i}{2}E^K_S$ for the chiral-odd part.
The subscript $S$ and the superscript $K$ denote different spin and transverse momentum dependence respectively.
The contributions from the quark-gluon-quark correlator can then be replaced by using these relations.

\section{Kinematic analysis of $e^+e^-\to V\pi X$}\label{sec:Kinematics}
\begin{figure}[!ht]
\centering \includegraphics[width=0.24\textwidth]{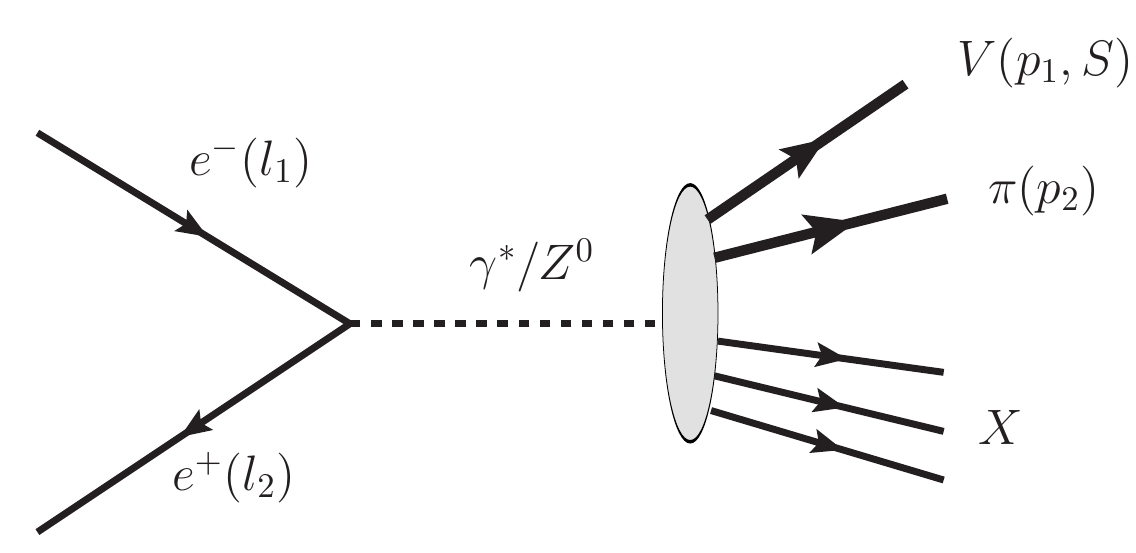}
\caption{Illustrating diagram for $e^+e^-\to V\pi X$.}
\label{fig:ff1}
\end{figure}

The best process for studying FFs is electron positron annihilation, because there is no initial PDFs involved.
Inclusive hadron production process is the simplest case to study one dimensional FFs. 
In order to study transverse momentum dependence, we must go to semi-inclusive process, 
see e.g.,~\cite{Wei:2013csa,Wei:2014pma} for these discussions.
To access both chiral-even and chiral-odd as well as the tensor polarization dependent FFs, 
the two-particle semi-inclusive $e^+e^-$-annihilation with vector meson production process as illustrated in Fig.~\ref{fig:ff1} is the best choice.

We consider the general weak interaction process $e^+e^-\to Z^0\to V\pi X$.
The differential cross section is given by
\begin{align}
\frac{2E_1E_2 d\sigma}{d^3p_1 d^3p_2} = \frac{\alpha^2 \chi}{sQ^4} L_{\mu\nu}(l_1,l_2) W^{\mu\nu}(q,p_1,S,p_2),
\label{eq:CS}
\end{align}
where $L_{\mu\nu}(l_1,l_2)$ is the well-known leptonic tensor,
and the hadronic tensors is defined as
\begin{align}
W_{\mu\nu} &(q,p_1,S,p_2) = \frac{1}{(2\pi)^4}  \sum_X (2\pi)^4 \delta^4 (q-p_1-p_2- p_X) \nonumber\\
& ~~~\times \langle 0| J_\nu (0) |p_1,S,p_2,X\rangle \langle p_1,S,p_2,X |J_\mu (0)|0\rangle.
\label{eq:HadronicTensor}
\end{align}
The first step we should do is to give a general kinematic analysis for this reaction, 
which is to construct the general form of the hadronic tensor,
and then calculate the cross section expression in terms of the structure functions.

\subsection{General form of the hadronic tensor and cross section} \label{subsec:WuvCS}
The hadronic tensor in Eq.~(\ref{eq:HadronicTensor}) satisfies constraints imposed by Hermiticity, current conservation.
We emphasize that if we consider annihilation via virtual photon, then parity conservation is also a constraint,
but it is not valid here for weak interaction. 
We need to construct the basic Lorentz tensors satisfy these constraints using the kinematic variables, 
and then decompose $W^{\mu\nu}$ under these basic tensors.

A systematic analysis of this decomposition for double spin-1/2 hadrons production 
has been presented in \cite{Pitonyak:2013dsu}. 
Here, we extend the analysis to include spin-1 hadron production,
and show the interesting similarities between the unpolarized and polarization dependent parts.

For the unpolarized part, the basic Lorentz tensors can only be constructed from $p_1$, $p_2$ and $q$, 
after some careful analysis, we found that the independent basic tensors can be written as
\begin{align}
h_{Ui}^{S\mu\nu}=& \Bigl\{ g^{\mu\nu}-\frac{q^\mu q^\nu}{q^2}, ~ p_{1q}^{\mu} p_{1q}^{\nu}, ~ p_{1q}^{\{\mu} p_{2q}^{\nu\}}, ~ p_{2q}^{\mu} p_{2q}^{\nu}  \Bigr\}, \\
\tilde h_{Ui}^{S\mu\nu}=& \Bigl\{ \varepsilon^{\{\mu qp_1p_2}\bigl(p_{1q}, ~ p_{2q}\bigr)^{\nu\}} \Bigr\}, \\
h_U^{A\mu\nu}=& p_{1q}^{[\mu} p_{2q}^{\nu]}, \\
\tilde h_{Ui}^{A\mu\nu}=& \Bigl\{ \varepsilon^{\mu\nu qp_1},  ~\varepsilon^{\mu\nu qp_2} \Bigr\}, 
\label{eq:hU}
\end{align}
where $h$'s are parity conserved ($P$-even) tensors,
while $\tilde h$'s are parity violated ($P$-odd) tensors.
The superscript $S$ or $A$ denotes symmetric or anti-symmetric under exchange of Lorentz index, 
the subscript $U$ denotes the unpolarized part.
A 4-momentum $p$ with a subscript $q$ denotes $p_q \equiv p-q(p\cdot q)/q^2$ satisfying $p_q\cdot q=0$,
so the current conservation is manifest. 
Some shorthand notations are used to make the expressions more concise, 
e.g., $a^{\{\mu} (b,c)^{\nu\}}$ means $a^{\{\mu}b^{\nu\}}$ and $a^{\{\mu}c^{\nu\}}$.
We see that there are 9 basic tensors in the unpolarized case. 

For the vector polarization dependent part, we have
\begin{align}
\begin{Bmatrix}
h_{Vi}^{S\mu\nu}\\ 
\tilde h_{Vi}^{S\mu\nu}\\ 
h_{Vi}^{A\mu\nu}\\ 
\tilde h_{Vi}^{A\mu\nu}
\end{Bmatrix}
=
\begin{Bmatrix}
\bigl[ (q\cdot S),(p_2\cdot S)\bigr] 
\begin{bmatrix}
\tilde h_{Ui}^{S\mu\nu}\\ 
h_{Ui}^{S\mu\nu}\\ 
\tilde h_{Ui}^{A\mu\nu}\\ 
h_{U}^{A\mu\nu}
\end{bmatrix},
&\varepsilon^{Sqp_1p_2}
\begin{bmatrix}
h_{Uj}^{S\mu\nu}\\ 
\tilde h_{Uj}^{S\mu\nu}\\ 
h_U^{A\mu\nu}\\ 
\tilde h_{Uj}^{A\mu\nu}
\end{bmatrix}
\end{Bmatrix}
\label{eq:hS}
\end{align}
There are in total 27 $S$-dependent basic tensors, corresponding to 3 independent vector polarization modes.
We see that they share exactly the same structures as the unpolarized part.
We only need to multiply in front the spin dependent scalar or pseudoscalar.
 
For the tensor polarized part, they also have the same structures as the unpolarized part.
For the $S_{LL}$-dependent basic tensors, because $S_{LL}$ is a scalar, 
they are just given by the unpolarized tensors multiplied by $S_{LL}$.
Therefore, we have 9 such tensors in this case. 
The $S_{LT}$-dependent part is given by
\begin{align}
\begin{Bmatrix}
h_{LTi}^{S\mu\nu}\\ 
\tilde h_{LTi}^{S\mu\nu}\\ 
h_{LTi}^{A\mu\nu}\\ 
\tilde h_{LTi}^{A\mu\nu}
\end{Bmatrix}
=
\begin{Bmatrix}
(p_2\cdot S_{LT})
\begin{bmatrix}
h_{Ui}^{S\mu\nu}\\ 
\tilde h_{Ui}^{S\mu\nu}\\ 
h_{U}^{A\mu\nu}\\ 
\tilde h_{Ui}^{A\mu\nu}
\end{bmatrix},
&\varepsilon^{S_{LT}qp_1p_2}
\begin{bmatrix}
\tilde h_{Uj}^{S\mu\nu}\\ 
h_{Uj}^{S\mu\nu}\\ 
\tilde h_{Uj}^{A\mu\nu}\\ 
h_{U}^{A\mu\nu}
\end{bmatrix}
\end{Bmatrix}
\label{eq:hSLT}
\end{align}
There are 18 such tensors in total, corresponding to the two independent $S_{LT}$-components.
The $S_{TT}$-dependent part takes exactly the same form as the $S_{LT}$-dependent part.
They are given by changing $S_{LT}^\alpha$ to $S_{TT}^{p_2\alpha}$ in Eq.~(\ref{eq:hSLT}).
So we have in total 81 basic Lorentz tensors for $W_{\mu\nu}(q,p_1,S,p_2)$, 41 of them are $P$-even and 40 are $P$-odd. 

The hadronic tensor is then expressed as a sum of all these basic Lorentz tensors multiplied by corresponding coefficients, i.e.,
$W^{\mu\nu} = W^{S\mu\nu} + iW^{A\mu\nu}$, 
with $W^{S/A\mu\nu} = \sum_{\sigma,i} W^{S/A}_{\sigma i} h_{\sigma i}^{S/A\mu\nu} + \sum_{\sigma,j} \tilde W^{S/A}_{\sigma j} \tilde h_{\sigma j}^{S/A\mu\nu}$.
The subscript $\sigma$ denotes $U$, $V$, $LL$, $LT$ and $TT$ for different polarizations,
and all the coefficients are real scalar functions of the Lorentz scalars 
$s=q^2$, $\xi_1=2q\cdot p_1/q^2$, $\xi_2=2q\cdot p_2/q^2$ and $\xi_{12}=s_{12}/s=(p_1+p_2)^2/s$. 

\begin{figure}[!ht]
\centering \includegraphics[width=0.3\textwidth]{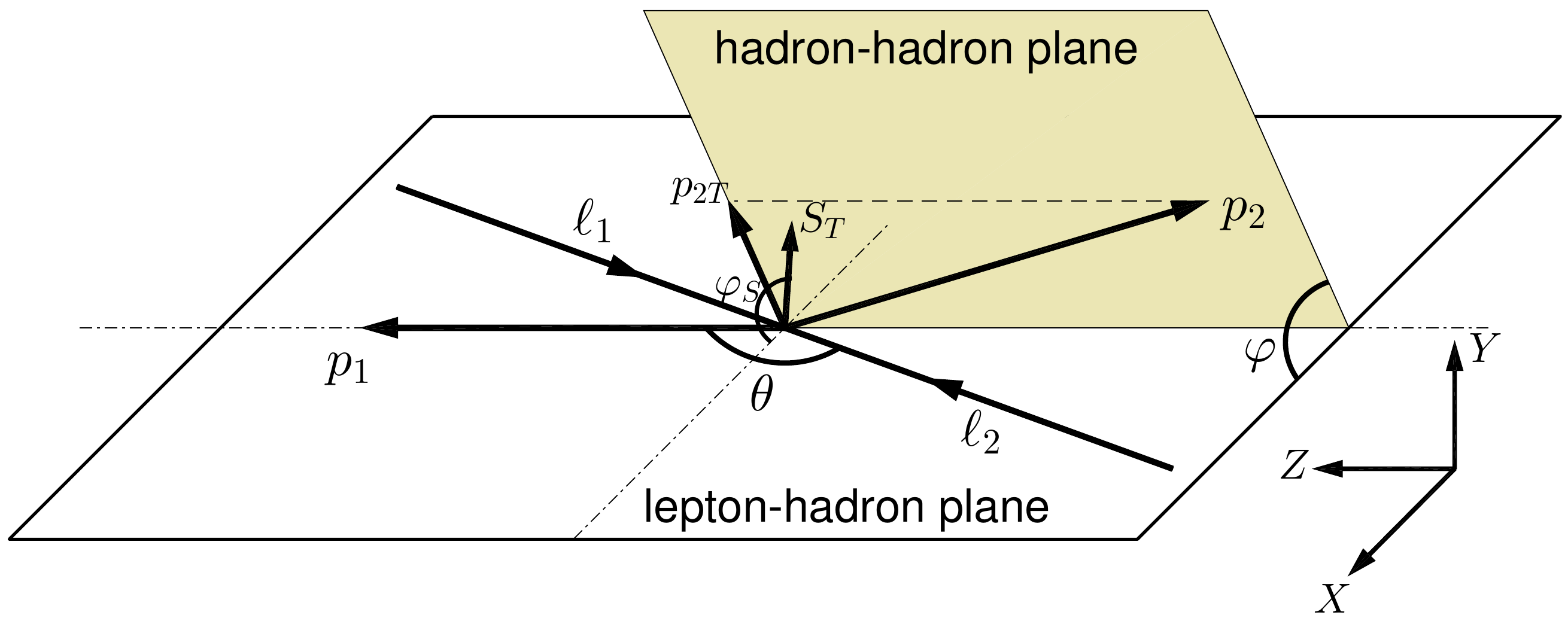}
\caption{(color on line). Illustrating diagram of the Helicity-GJ frame.}
\label{fig:frame}
\end{figure}

\vspace{2mm}
Substitute $W^{\mu\nu}(q,p_1,S,p_2)$ into Eq.~(\ref{eq:CS}), we will get the cross section in Lorentz invariant form.
Here, we write down the expressions in a special reference frame.
We choose the helicity frame of $V$ ($\vec p_1$ along positive $z$-direction) which is suitable for studying the vector meson polarization.
We also choose center of mass frame of the leptons,
with the lepton-hadron plane as $Oxz$ plane, as Fig.~\ref{fig:frame} shows.
This is a particular Gottfried-Jackson frame~\cite{Gottfried:1964nx} which we will refer to as ``Helicity-GJ frame''. 

In this frame, the cross section can be expressed as
\begin{align}
\frac{E_1E_2d\sigma}{d^3p_1d^3p_2} =& \frac{\alpha^2\chi}{2s^2} \Bigl[ ({\cal F}_U + \tilde{\cal F}_U) + \lambda ({\cal F}_L + \tilde{\cal F}_L) \nonumber\\
+& |S_T| ({\cal F}_T + \tilde{\cal F}_T) + S_{LL} ({\cal F}_{LL} + \tilde{\cal F}_{LL}) \nonumber\\
+& |S_{LT}| ({\cal F}_{LT} + \tilde{\cal F}_{LT}) + |S_{TT}| ({\cal F}_{TT} + \tilde{\cal F}_{TT})  \Bigr],
\end{align}
For the unpolarized part, we have
\begin{align}
{\cal F}_U =& (1+\cos^2\theta) F_{1U}+  \sin^2\theta F_{2U} + \cos\theta F_{3U} \nonumber\\
&+ \cos\varphi \bigl[ \sin\theta F_{1U}^{\cos\varphi} +  \sin2\theta F_{2U}^{\cos\varphi}\bigr] \nonumber\\
&+\cos2\varphi \sin^2\theta F_U^{\cos2\varphi}, \\
\tilde{\cal F}_U=& \sin\varphi  \bigl[ \sin\theta \tilde F_{1U}^{\sin\varphi} + \sin2\theta \tilde F_{2U}^{\sin\varphi} \bigr] \nonumber\\
&+ \sin2\varphi\sin^2\theta \tilde F_U^{\sin2\varphi},
\end{align} 
$F_U$'s and $\tilde F_U$'s are called structure functions and are all scalar functions of $s$, $\xi_1, \xi_2$ and $p_{2T}^2$.
Among them, the 6 $F_{U}$'s correspond to parity conserving terms of the cross section, 
while the other 3 $\tilde F_{U}$'s are parity odd part. 

For the polarized part,
since $\lambda$ and $S_{LL}$ are just pseudoscalar and scalar, 
the structure functions related to them have one to one correspondence to the unpolarized part, i.e.,
\begin{align}
\Bigl\{ {\cal F}_L, ~ \tilde{\cal F}_L \Bigr\} \Leftrightarrow \Bigl\{ \tilde{\cal F}_U, ~ {\cal F}_U \Bigr\}, \quad
\Bigl\{ {\cal F}_{LL}, ~ \tilde{\cal F}_{LL} \Bigr\} \Leftrightarrow \Bigl\{ {\cal F}_U, ~ \tilde{\cal F}_U \Bigr\}.
\end{align}
The transverse polarized parts also share the similar structure as the unpolarized part,
the only difference is the transverse polarization angle dependence. 
We just show here the parity conserved $S_T$ dependent part ${\cal F}_T$ for an example, i.e.,
\begin{align}
{\cal F}_T &=
\sin\varphi_S \bigl[ \sin\theta F_{1T}^{\sin\varphi_S} + \sin2\theta F_{2T}^{\sin\varphi_S} \bigr] \nonumber \\
& + \sin(\varphi_S + \varphi) \sin^2\theta F_{T}^{\sin(\varphi_S + \varphi)} \nonumber\\
& + \sin(\varphi_S - \varphi) \bigl[ (1+\cos^2\theta) F_{1T}^{\sin(\varphi_S - \varphi)} \nonumber\\
&\qquad\qquad\qquad + \sin^2\theta F_{2T}^{\sin(\varphi_S - \varphi)} + \cos\theta F_{3T}^{\sin(\varphi_S - \varphi)} \bigr] \nonumber\\
& + \sin(\varphi_S - 2\varphi) \bigl[ \sin\theta F_{1T}^{\sin(\varphi_S - 2\varphi)} + \sin2\theta F_{2T}^{\sin(\varphi_S - 2\varphi)} \bigr] \nonumber \\
& + \sin(\varphi_S - 3\varphi) \sin^2\theta F_{T}^{\sin(\varphi_S - 3\varphi)}.\label{eq:FT}
\end{align}
We see the additional polarization angle dependence due to terms of $\varepsilon^{Sqp_1p_2}$ and $(p_2\cdot S)$.

\subsection{The azimuthal asymmetries and hadron polarizations}
Once we get the cross section results, 
we can derive physical observables such as hadron azimuthal asymmetries and polarizations straightforwardly. 

For measuring azimuthal asymmetries, people often sum up the hadron polarizations, 
which correspond to the unpolarized hadron production case.
In this case we have four azimuthal asymmetries due to the $\varphi$ dependent terms in the cross section, i.e.,
\begin{align}
&\langle \cos\varphi\rangle_U =(\sin\theta F_{1U}^{\cos\varphi} + \sin2\theta F_{2U}^{\cos\varphi})/{2F_{Ut}}, \label{eq:Acos}\\
&\langle \cos2\varphi\rangle_U={\sin^2\theta F_{U}^{\cos2\varphi}}/{2F_{Ut}}, \label{eq:Acos2}\\
&\langle \sin\varphi\rangle_U =(\sin\theta \tilde F_{1U}^{\sin\varphi} + \sin2\theta \tilde F_{2U}^{\sin\varphi})/{2F_{Ut}}, \label{eq:Asin}\\
&\langle \sin2\varphi\rangle_U={\sin^2\theta \tilde F_{U}^{\sin2\varphi}}/{2F_{Ut}}, \label{eq:Asin2}
\end{align}
where $F_{Ut}$ denotes the sum of the structure functions averaged over $\varphi$, i.e, 
$F_{Ut} = (1+\cos^2\theta) F_{1U}+  \sin^2\theta F_{2U} + \cos\theta F_{3U}$.
We point out that the cosine-asymmetries correspond to parity conserving part,
while the sine-asymmetries correspond to parity violating part which only appear in weak interaction process.

For measuring the polarization of the vector meson, we often average over the hadron azimuthal angle.
In this case, we have, for the longitudinal polarization
\begin{align}
&\langle \lambda \rangle=\frac{2}{3F_{Ut}} \Bigl[ (1+\cos^2\theta) \tilde F_{1L}+  \sin^2\theta \tilde F_{2L} + \cos\theta \tilde F_{3L}\Bigr], \label{eq:Alambda}\\
&\langle S_{LL}\rangle= \frac{1}{2F_{Ut}} \Bigl[ (1+\cos^2\theta) F_{1LL}+  \sin^2\theta F_{2LL} + \cos\theta F_{3LL} \Bigr]. \label{eq:ASll}
\end{align}
We see that they correspond to parity violated and conserved structures respectively.
For the transverse polarizations, we can measure the components with respect to the lepton-hadron plane,
but a more convenient choice is to measure the components normal or tangent to the hadron-hadron plane,
i.e., the two transverse directions defined as 
$\vec e_n=\vec p_1\times\vec p_2/ |\vec p_1\times\vec p_2|$ and 
$\vec e_t=\vec p_{2T}/|\vec p_{2T}|$, 
and then average the value over azimuthal angle $\varphi$.
In this case, the results are give by
\begin{align}
\langle S_{T}^n\rangle=& \frac{2}{3F_{Ut}} \Bigl[ (1+\cos^2\theta) F_{1T}^{\sin(\varphi_S - \varphi)} \nonumber\\
&+ \sin^2\theta F_{2T}^{\sin(\varphi_S - \varphi)} + \cos\theta F_{3T}^{\sin(\varphi_S - \varphi)} \Bigr], \label{eq:AStn}\\
\langle S_{T}^t\rangle=&\frac{2}{3F_{Ut}} \Bigl[ (1+\cos^2\theta) \tilde F_{1T}^{\cos(\varphi_S - \varphi)} \nonumber\\
& + \sin^2\theta \tilde F_{2T}^{\cos(\varphi_S - \varphi)} + \cos\theta \tilde F_{3T}^{\cos(\varphi_S - \varphi)}\Bigr], \label{eq:AStt}\\
\langle S_{LT}^n\rangle=& \frac{2}{3F_{Ut}} \Bigl[ (1+\cos^2\theta) \tilde F_{1LT}^{\sin(\varphi_{LT} - \varphi)} \nonumber\\
&+ \sin^2\theta \tilde F_{2LT}^{\sin(\varphi_{LT} - \varphi)} + \cos\theta \tilde F_{3LT}^{\sin(\varphi_{LT} - \varphi)} \Bigr],\label{eq:ASltn}\\
\langle S_{LT}^t\rangle=&\frac{2}{3F_{Ut}}\Bigl[ (1+\cos^2\theta) F_{1LT}^{\cos(\varphi_{LT} - \varphi)} \nonumber\\
& + \sin^2\theta F_{2LT}^{\cos(\varphi_{LT} - \varphi)} + \cos\theta F_{3LT}^{\cos(\varphi_{LT} - \varphi)} \Bigr], \label{eq:ASltt}\\
\langle S_{TT}^{nn}\rangle=& \frac{-2}{3F_{Ut}} \Bigl[ (1+\cos^2\theta) F_{1TT}^{\cos(2\varphi_{TT} - 2\varphi)}  \nonumber\\
& + \sin^2\theta F_{2TT}^{\cos(2\varphi_{TT} - 2\varphi)} + \cos\theta F_{3TT}^{\cos(2\varphi_{TT} - 2\varphi)} \Bigr], \label{eq:ASttnn}\\
\langle S_{TT}^{nt}\rangle=&\frac{2}{3F_{Ut}} \Bigl[ (1+\cos^2\theta) \tilde F_{1TT}^{\sin(2\varphi_{TT} - 2\varphi)}  \nonumber\\
& + \sin^2\theta \tilde F_{2TT}^{\sin(2\varphi_{TT} - 2\varphi)} + \cos\theta \tilde F_{3TT}^{\sin(2\varphi_{TT} - 2\varphi)} \Bigr]. \label{eq:ASttnt}
\end{align}
It is interesting to see that the transverse polarizations just correspond to the structure functions of 
$\cos(\varphi_\sigma-\varphi)$- or $\sin(\varphi_\sigma-\varphi)$-terms.
We also see that in this case $\langle S_{T}^n\rangle$, $\langle S_{LT}^t\rangle$ and $\langle S_{TT}^{nn}\rangle$ are parity conserved, 
while the other three are parity violated. 

From the above kinematic analysis results, we see that 
we can study the azimuthal asymmetries in the unpolarized case, 
or study the longitudinal hadron polarization in the helicity frame and transverse polarizations w.r.t.
the hadron-hadron plane averaged over the azimuthal angle $\varphi$.
By measuring these quantities experimentally we can obtain the information of the corresponding structure functions.

\section{Parton model results}\label{sec:PartonModel}
\subsection{Structure functions results in terms of TMD FFs}
\begin{figure}[!ht]
  \centering 
  \subfigure[]{ 
    \label{subfig:twist-2} 
    \includegraphics[width=0.17\textwidth, height=2cm]{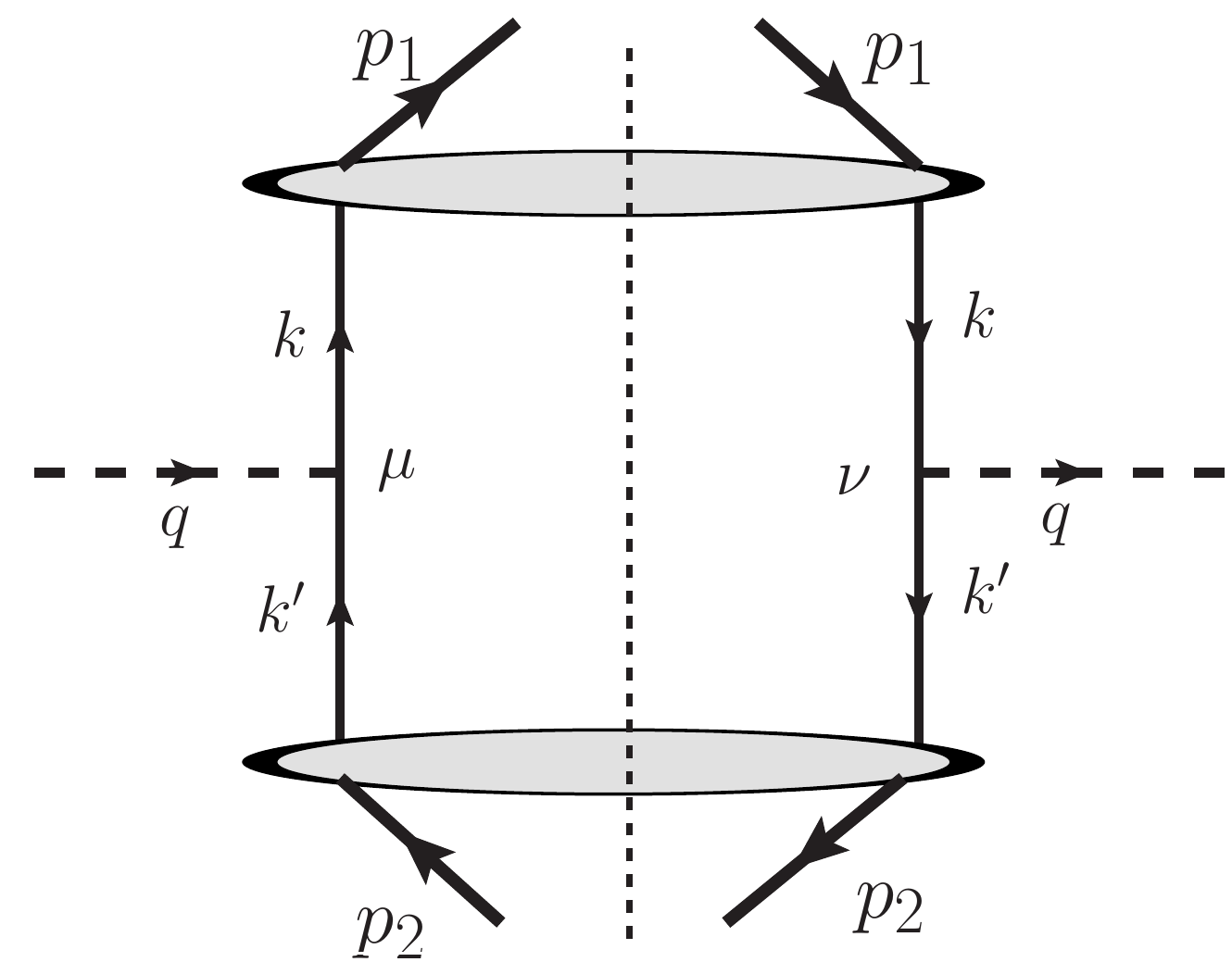}} 
  \hspace{5mm} 
  \subfigure[]{ 
    \label{subfig:twist-3} 
    \includegraphics[width=0.17\textwidth, height=2cm]{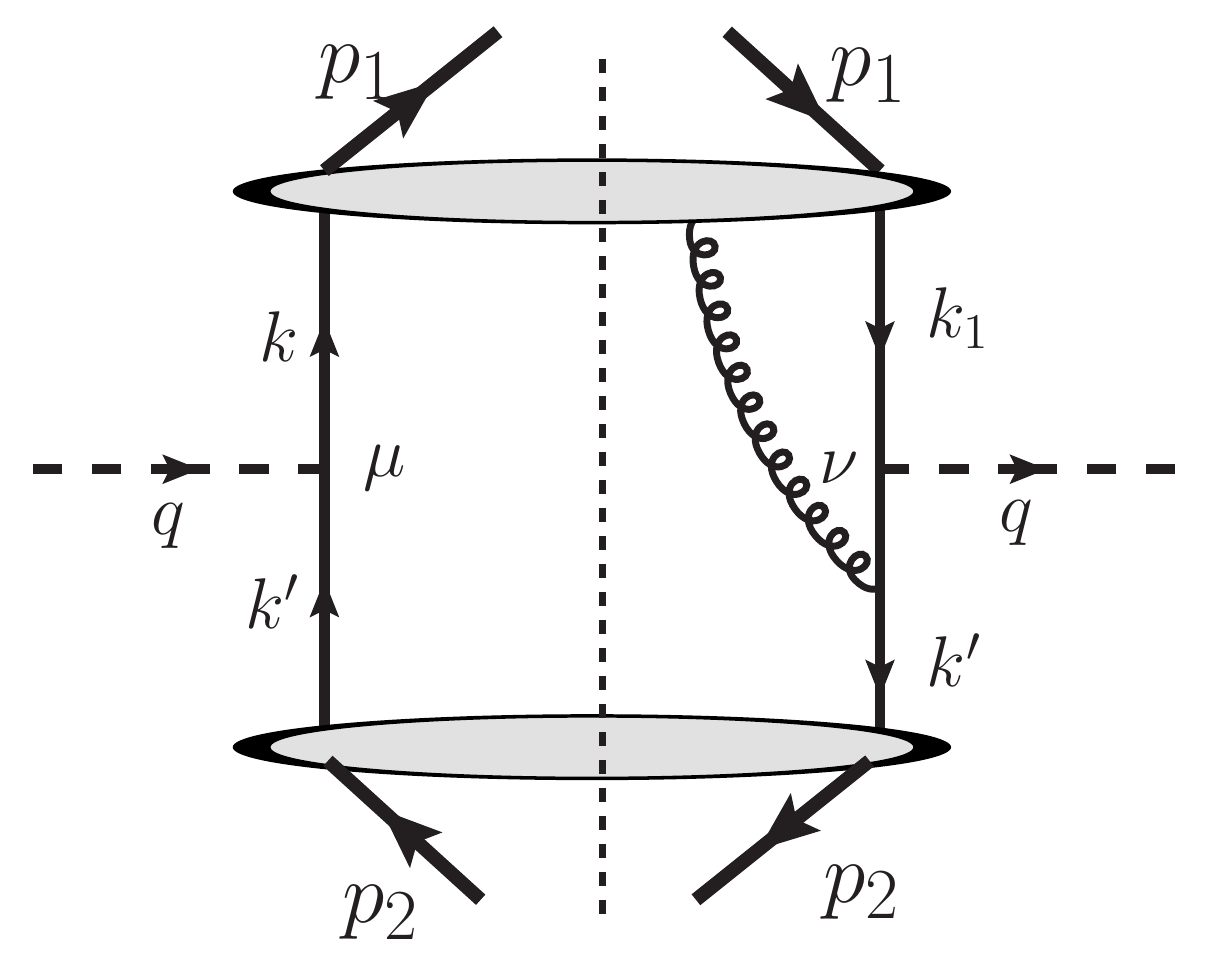}}
  \caption{Feynman diagrams for the hadronic tensor. (a) contributes at leading and higher twist; (b) contributes at twist-3 and higher twist.} 
  \label{fig:PartonModel} 
\end{figure}

The above kinematic analysis results are model independent.
We can also calculate the hadronic tensor using parton model so that the differential cross section is expressed using FFs.
We make the calculations at leading order in pQCD but up to twist-3 level. 
Similar calculations for double polarized spin-1/2 hadron production has been investigated e.g. in~\cite{Boer:1997mf}.
To this end, we need to consider the Feynman diagrams shown in Fig.~\ref{fig:PartonModel}.
The leading twist contribution comes from Fig.~\ref{subfig:twist-2}.
Twist-3 contribution comes from both Fig.~\ref{subfig:twist-2} and Fig.~\ref{subfig:twist-3}
There are also three similar diagrams as Fig.~\ref{subfig:twist-3} with the gluon attached to the quark line 
and the complex conjugate contribution, we do not show them here.

Corresponding to these Feynman diagrams, the hadronic tensor is written as 
$W_{\mu\nu}=\tilde W_{\mu\nu}^{(0)} +\tilde W_{\mu\nu}^{(1)}-\Delta\tilde W_{\mu\nu}^{(0)}$. 
For the leading twist contribution $\tilde W_{\mu\nu}^{(0)}$ from Fig.~\ref{subfig:twist-2}, we have
\begin{align}
\tilde W_{\mu\nu}^{(0)}=&\frac{1}{p_1^+ p_2^-} \int \frac{d^2k_\perp}{(2\pi)^2} \frac{d^2k^\prime_\perp}{(2\pi)^2} \delta^2(k_\perp + k^\prime_\perp - q_\perp) \nonumber\\ 
&~\times {\rm Tr} \bigl[ \hat \Xi^{(0)}(z_1,k_\perp,p_1,S) \Gamma_\mu \hat {\bar\Xi}^{(0)}(z_2,k^\prime_\perp,p_2) \Gamma_\nu \bigr].  \label{eq:HT0}
\end{align}
Corresponding to Fig.~\ref{subfig:twist-3}, we have the twist-3 contribution
\begin{align}
\tilde W_{\mu\nu}^{(1b)} =& \frac{-1}{\sqrt{2}Qp_1^+ p_2^-}  \int \frac{d^2k_\perp}{(2\pi)^2} \frac{d^2k^\prime_\perp}{(2\pi)^2} \delta^2(k_\perp + k^\prime_\perp - q_\perp)  \nonumber\\
 \times & {\rm Tr} \bigl[ \Gamma_\mu \hat {\bar\Xi}^{(0)}(z_2,k^\prime_\perp,p_2) \gamma_\rho \slashed{\bar n} \Gamma_\nu \hat \Xi^{(1)\rho}(z_1,k_\perp,p_1,S) \bigr],  \label{eq:HT1}\\
\Delta\tilde W_{\mu\nu}^{(0b)}&=
\frac{1}{\sqrt{2}Q p_1^+ p_2^-} \int \frac{d^2k_\perp}{(2\pi)^2} \frac{d^2k^\prime_\perp}{(2\pi)^2} \delta^2(k_\perp + k^\prime_\perp - q_\perp) \nonumber\\
\times  k_\perp^\rho & {\rm Tr} \bigl[ \Gamma_\mu \hat {\bar\Xi}^{(0)}(z_2,k^\prime_\perp,p_2) \gamma_\rho \slashed{\bar n} \Gamma_\nu \hat \Xi^{(0)}(z_1,k_\perp,p_1,S) \bigr].\label{eq:dHT1}
\end{align}
We see that the hadronic tensor is expressed by the quark-quark and quark-gluon-quark correlator.
Substitute the correlator's decomposition results and carry out the trace, 
we get the hadronic tensor in terms of TMD FFs.
Then making Lorentz contraction with the leptonic tensor, we obtain the cross section and
by comparing with the kinematic analysis results, 
we will get the structure functions results in terms of the TMD FFs. 
Here, we directly show the structure functions results which have leading twist correspondence.

For the unpolarized part, we have
\begin{align}
& F_{1U1} ={2c_1^e c_1^q} \mathcal{C} [ D_1 \bar D_1 ], \label{eq:F1U1} \\
& F_{3U1} = {4c_3^e c_3^q} \mathcal{C} [ D_1 \bar D_1 ],  \label{eq:F3U1} \\
& F_{U1}^{\cos 2\varphi} = -{8c_1^e c_2^q} \mathcal{C} [ w_{hh} H_1^\perp \bar H_1^\perp ], \label{eq:FU1cos2phi}
\end{align}
where we have defined the convolution of the TMD FFs, e.g.,
\begin{align}
\mathcal{C} [ w D \bar D ] = \frac{1}{z_1z_2}\int & \frac{d^2k_\perp}{(2\pi)^2} \frac{d^2k^\prime_\perp}{(2\pi)^2} \delta^2(k_\perp +k_\perp^\prime - q_\perp)\nonumber\\
& \times w(k_\perp,k_\perp^\prime) D(z_1,k_\perp) \bar D(z_2,k_\perp^\prime),
\end{align}
and $w$ is a dimensionless scalar weight function of $k_\perp$ and $k'_\perp$.
Here, we introduce a second digital ``1" in the structure function's subscript to denote it has leading twist correspondence.
We note in particular that among the azimuthal angle dependent terms, 
only the $\cos2\varphi$ term has twist-2 contribution due to Collins function~\cite{Collins:1992kk}.

For the longitudinal vector polarization dependent part, 
we have also three nonzero structure functions at twist-2 and they are very much similar to the unpolarized part, i.e.,
\begin{align}
& \tilde F_{1L1} = -{2c_1^e c_3^q} \mathcal{C} [ G_{1L} \bar D_1 ], \label{eq:tF1L1} \\
& \tilde F_{3L1} = -{4c_3^e c_1^q} \mathcal{C} [ G_{1L} \bar D_1 ], \label{eq:tF3L1} \\
& F_{L1}^{\sin2\varphi} = -{8c_1^e c_2^q} \mathcal{C} [ w_{hh} H_{1L}^\perp \bar H_1^\perp ]. \label{eq:FLsin2phi}
\end{align}
For the transverse vector polarization dependent part, we have
\begin{align}
& F_{1T1}^{\sin(\varphi_S-\varphi)} ={2c_1^e c_1^q} \mathcal{C} [ w_1 D_{1T}^\perp \bar D_1 ], \label{eq:F1T1} \\
& F_{3T1}^{\sin(\varphi_S-\varphi)} = {4c_3^e c_3^q} \mathcal{C} [ w_1 D_{1T}^\perp \bar D_1 ], \label{eq:F3T1}\\
& \tilde F_{1T1}^{\cos(\varphi_S-\varphi)} ={2c_1^e c_3^q} \mathcal{C} [ w_1 G_{1T}^\perp \bar D_1 ], \label{eq:tF1T1} \\
& \tilde F_{3T1}^{\cos(\varphi_S-\varphi)} = {4c_3^e c_1^q} \mathcal{C} [ w_1 G_{1T}^\perp \bar D_1 ], \label{eq:tF3T1} \\
& F_{T1}^{\sin(\varphi_S+\varphi)} = -{8c_1^e c_2^q} \mathcal{C} [ \bar w_1 {\mathcal H}_{1T}^{\perp} \bar H_1^\perp ], \\
& F_{T1}^{\sin(\varphi_S-3\varphi)} = -{8c_1^e c_2^q} \mathcal{C} [ w_{hh}^t H_{1T}^\perp \bar H_1^\perp ].
\end{align}
We see that there are 6 non-zero transverse polarization dependent structure functions at twist-2, 
4 of them are parity conserved and the other 2 are parity violated. 

For the tensor polarization dependent part,
first of all, the $S_{LL}$-dependent part is exactly the same as the unpolarized part. 
We only need to change the corresponding subscript to $LL$ to get the results.
The $S_{LT}$-dependent part is very much similar to the $S_T$-part. 
The main difference comes from parity, since $S_{LT}$ is a vector.
The 6 non-zeros are given by
\begin{align}
& F_{1LT1}^{\cos(\varphi_{LT}-\varphi)} = -{2c_1^e c_1^q} \mathcal{C} [ w_1 D_{1LT}^\perp \bar D_1 ], \\
& F_{3LT1}^{\cos(\varphi_{LT}-\varphi)} = -{4c_3^e c_3^q} \mathcal{C} [ w_1 D_{1LT}^\perp \bar D_1 ], \\
& \tilde F_{1LT1}^{\sin(\varphi_{LT}-\varphi)} = -{2c_1^e c_3^q} \mathcal{C} [ w_1 G_{1LT}^\perp \bar D_1 ], \\
& \tilde F_{3LT1}^{\sin(\varphi_{LT}-\varphi)} = -{4c_3^e c_1^q} \mathcal{C} [ w_1 G_{1LT}^\perp \bar D_1 ], \\
& F_{LT1}^{\cos(\varphi_{LT}+\varphi)} = -{8c_1^e c_2^q} \mathcal{C} [ \bar w_1 {\cal H}_{1LT}^\perp \bar H_1^\perp ], \\
& F_{LT1}^{\cos(\varphi_{LT}-3\varphi)} = {8c_1^e c_2^q} \mathcal{C} [ w_{hh}^{t}H_{1LT}^\perp \bar H_1^\perp ]. 
\end{align}
The $S_{TT}$-dependent part is similar to the $S_{LT}$-part but the weights are different, the results are
\begin{align}
& F_{1TT1}^{\cos(2\varphi_{TT}-2\varphi)} = {2c_1^e c_1^q} \mathcal{C} [ w_{dd}^{tt} D_{1TT}^\perp \bar D_1 ], \\
& F_{3TT1}^{\cos(2\varphi_{TT}-2\varphi)} = {4c_3^e c_3^q} \mathcal{C} [ w_{dd}^{tt} D_{1TT}^\perp \bar D_1 ], \\
& \tilde F_{1TT1}^{\sin(2\varphi_{TT}-2\varphi)} = {2c_1^e c_3^q} \mathcal{C} [ w_{dd}^{tt} G_{1TT}^\perp \bar D_1 ], \\
& \tilde F_{3TT1}^{\sin(2\varphi_{TT}-2\varphi)} = {4c_3^e c_1^q} \mathcal{C} [ w_{dd}^{tt} G_{1TT}^\perp \bar D_1 ], \\
& F_{TT1}^{\cos(2\varphi_{TT}-4\varphi)} = -{4c_1^e c_2^q} \mathcal{C} [ w_{hh}^{tt} H_{1TT}^\perp \bar H_1^\perp ], \\
& F_{TT1}^{\cos2\varphi_{TT}} = {8c_1^e c_2^q} \mathcal{C} [ w_2 {H}_{1TT}^{\perp\prime}  \bar H_1^\perp ],
\end{align}

To summarize, we have in total 27 nonzero structure functions at leading twist level for $e^+e^- \to Z^0\to V\pi X$ process.
These leading twist structure functions are all related to the $1+\cos^2\theta$, $\cos\theta$ and $\sin^2\theta$ angular distribution,
which is a direct result corresponding to the parton level reaction $e^+e^-\to Z^0\to q\bar q$.
The twist-3 level results can be found in the appendix of Ref.~\cite{Chen:2016moq}.

\subsection{Azimuthal asymmetries and hadron polarizations}\label{subsec:APol}
From the  parton model results of the structure functions we can easily derive the hadron azimuthal asymmetries and polarizations in terms of TMD FFs.

At leading twist and for unpolarized case, there is only one azimuthal asymmetry as given by Eq.~(\ref{eq:Acos2}), i.e.,
(here and in the following, the summation over quark flavor is implicit for the numerator and denominator)
\begin{align}
\langle \cos2\varphi\rangle_{U}^{(0)} = - \frac{ C(y)  c_1^e c_2^q 
\mathcal{C} [ w_{hh} H_1^\perp \bar H_1^\perp ] }{  T_0^q(y) \mathcal{C} [ D_1 \bar D_1 ] },
\end{align}
this corresponds to the well-known Collins effect~\cite{Collins:1992kk} and is the result of $q\bar q$ transverse spin correlation.
At twist-3, we have another two azimuthal asymmetries, i.e., 
\begin{align}
&\langle\cos\varphi\rangle_U^{(1)}=  -F^{(1)}\biggl\{T_2^q(y) \Bigl(M_1 \mathcal{C} [ w_1 D^\perp z_2 \bar D_1]  + M_2  \mathcal{C} [ \bar w_1 z_1 D_1 \bar D^{\perp\prime}] \Bigr) \nonumber\\
&~~~~~ + T_4^q(y) \Bigl( M_1 \mathcal{C}[ \bar w_1 H z_2 \bar H_1^\perp] + M_2  \mathcal{C}[ w_1 z_1 H_1^\perp  \bar H^{\perp\prime}] \Bigr) \biggr\},\\ 
& \langle\sin\varphi\rangle_U^{(1)}=  F^{(1)}\biggl\{ T_3^q(y) \Bigl( M_1 \mathcal{C} [ w_1 G^\perp z_2 \bar D_1] - M_2 \mathcal{C} [ \bar w_1 z_1 D_1 \bar G^\perp]\Bigr) \nonumber\\
&~~~~~ + 2 c_3^e c_2^q \Bigl( M_1 \mathcal{C} [ \bar w_1 E z_2 \bar H_1^\perp] - M_2 \mathcal{C} [ w_1 z_1 H_1^\perp \bar E] \Bigr)  \biggr\},
\end{align}
where $F^{(1)} = 2D(y)/z_1z_2QT_0^q{\cal C}[D_1\bar D_1]$. The $\cos\varphi$ asymmetry 
correspond to the well-known Cahn effect in DIS process~\cite{Cahn:1978se}, 
and the $\sin\varphi$ asymmetry is the parity violating counterpart only appears in weak interaction process.

\vspace{2mm}
For hadron polarizations, we have the longitudinal polarization at twist-2 as
\begin{align}
&\langle \lambda\rangle^{(0)}
= \frac{2}{3} \frac{ P_q(y)T_0^q(y) \mathcal{C}[G_{1L} \bar D_1]}{ T_0^q(y) \mathcal{C}[D_1 \bar D_1]}, \label{eq:ALambdat2}\\
&\langle S_{LL}\rangle^{(0)}
= \frac{1}{2} \frac{ T_0^q(y) \mathcal{C} [ D_{1LL} \bar D_1 ]}{ T_0^q(y) \mathcal{C}[D_1 \bar D_1]}. \label{eq:ASllt2}
\end{align}
For the averaged transverse polarizations w.r.t. the hadron-hadron plane,  we have
\begin{align}
&\langle S_T^{n}\rangle^{(0)} = \frac{2}{3} \frac{ T_0^q(y) \mathcal{C} [ w_1 D_{1T}^\perp \bar D_1 ]}{ T_0^q(y) \mathcal{C}[D_1 \bar D_1]}, \label{eq:AStnt2}\\
&\langle S_T^{t}\rangle^{(0)} = -\frac{2}{3} \frac{  P_q(y) T_0^q(y)  \mathcal{C} [ w_1 G_{1T}^\perp \bar D_1 ]}{ T_0^q(y) \mathcal{C}[D_1 \bar D_1]},\label{eq:ASttt2}\\
&\langle S_{LT}^{n}\rangle^{(0)} = \frac{2}{3} \frac{ P_q(y) T_0^q(y)  \mathcal{C} [ w_1 G_{1LT}^\perp \bar D_1 ]}{ T_0^q(y) \mathcal{C}[D_1 \bar D_1]}, \label{eq:ASltnt2}\\
&\langle S_{LT}^{t}\rangle^{(0)} = -\frac{2}{3}\frac{ T_0^q(y)  \mathcal{C} [ w_1 D_{1LT}^\perp \bar D_1 ]}{ T_0^q(y) \mathcal{C}[D_1 \bar D_1]},\label{eq:ASlttt2}\\
&\langle S_{TT}^{nn}\rangle^{(0)} = -\frac{2}{3}\frac{ T_0^q(y)  \mathcal{C} [ w_{dd}^{tt} D_{1TT}^\perp \bar D_1 ]}{ T_0^q(y) \mathcal{C}[D_1 \bar D_1]}, \label{eq:ASttnnt2}\\
&\langle S_{TT}^{nt}\rangle^{(0)} = -\frac{2}{3}\frac{ P_q(y) T_0^q(y)  \mathcal{C} [ w_{dd}^{tt} G_{1TT}^\perp \bar D_1 ]}{ T_0^q(y) \mathcal{C}[D_1 \bar D_1]}.\label{eq:ASttntt2}
\end{align}
We see in particular that these polarizations at leading twist can be divided into two categories,
those depend on the quark polarization $P_q(y)$ i.e., 
$\langle \lambda\rangle$, $\langle S_T^{t}\rangle$, $\langle S_{LT}^{n}\rangle$ and $\langle S_{TT}^{nt}\rangle$, and are parity violated, 
while the others are independent of quark polarization and are parity conserved.

At twist-3 level, there is no transverse polarization w.r.t. the hadron-hadron plane.
However, four components i.e., $\langle S_T^x\rangle$, $\langle S_T^y\rangle$, $\langle S_{LT}^x\rangle$ and $\langle S_{LT}^y\rangle$
exist w.r.t. the lepton-hadron plane.
It is also interesting to see that these transverse components also exist in the inclusive process $e^+e^-\to Z^0\to VX$, we have 
\begin{align}
&\langle S_T^{x}\rangle^{(1)}_{in} = -F_{in}^{(1)} T_3^q(y) G_T,~~~~~\langle S_T^{y}\rangle^{(1)}_{in} = F_{in}^{(1)} T_2^q(y) D_T, \label{eq:AStt3}\\
&\langle S_{LT}^{x}\rangle^{(1)}_{in} = -F_{in}^{(1)} T_2^q(y) D_{LT},~\langle S_{LT}^{y}\rangle^{(1)}_{in} = F_{in}^{(1)} T_3^q(y) G_{LT}, \label{eq:ASltt3}
\end{align}
where $F_{in}^{(1)} = 8M_1D(y)/[3z_1Q T_0^q(y) D_1]$.
We would like to point out that only $\langle S_{LT}^{x}\rangle$ is both space reflection and time reversal even,
and has correspondence in inclusive DIS process. 
The other three are either space reflection or time reversal odd and should vanish in inclusive DIS process.

\section{Energy dependence of hadron polarizations}
We have seen from the above results that at leading twist the hadron longitudinal polarization
or helicity $\langle \lambda \rangle$ depends on quark polarization, 
while $\langle S_{LL} \rangle$ corresponds to vector meson spin alignment is independent of quark polarization.
Since the quark polarization produced in the electroweak process can be calculated and have strong energy dependence, 
we expect that the hadron polarizations for this two cases would have very different energy dependence.
Because there are some data on Lambda Hyperon longitudinal polarization\cite{LambdaPol} and $K^*$ meson spin alignment\cite{Rho00} at LEP,
we take them as examples and calculate their energy dependences.

\begin{figure}[!ht]
  \centering 
  \subfigure[]{ 
    \label{subfig:lambdapol} 
    \includegraphics[width=0.2\textwidth, height=28mm]{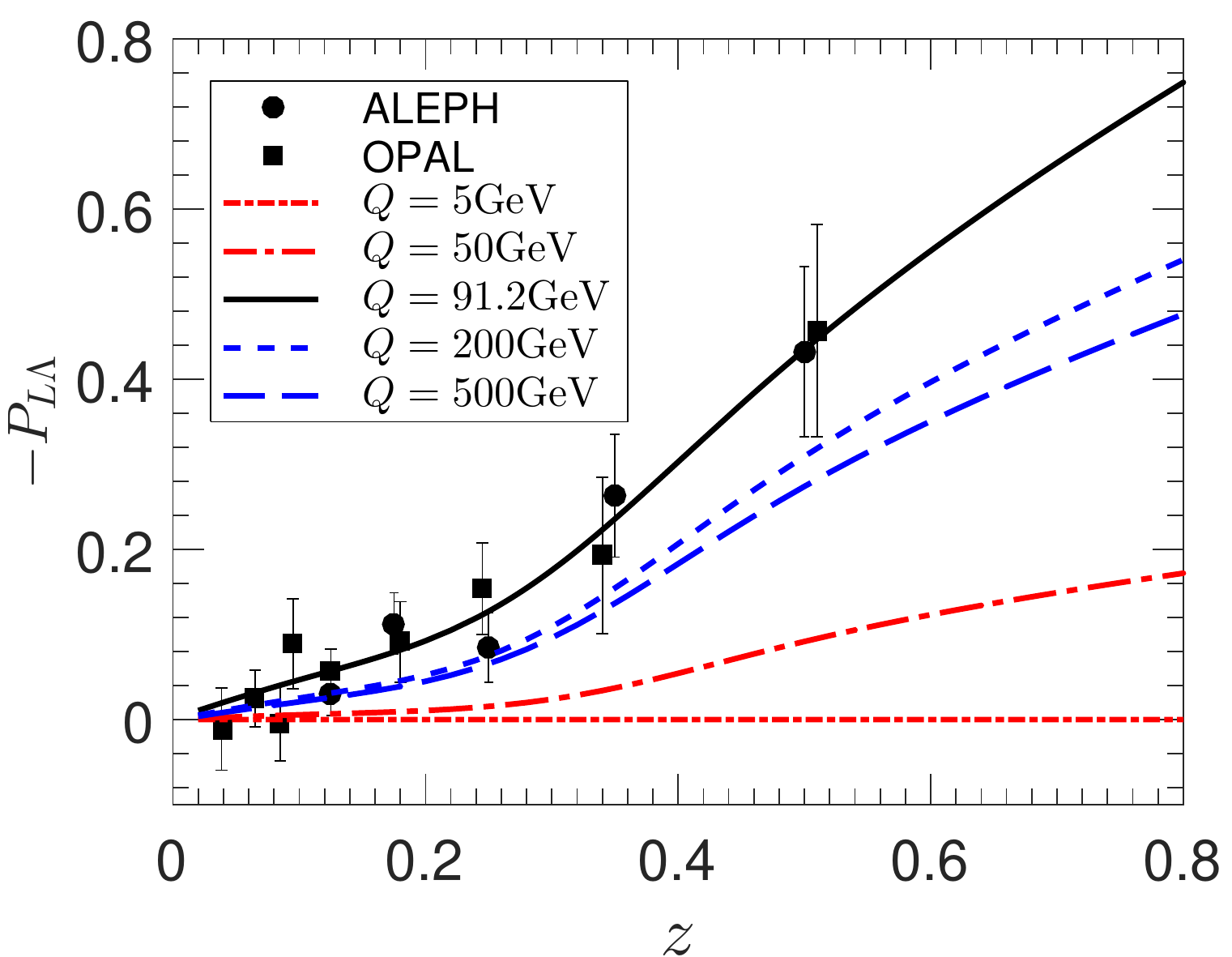}} 
  \hspace{5mm} 
  \subfigure[]{ 
    \label{subfig:rho00} 
    \includegraphics[width=0.2\textwidth, height=28mm]{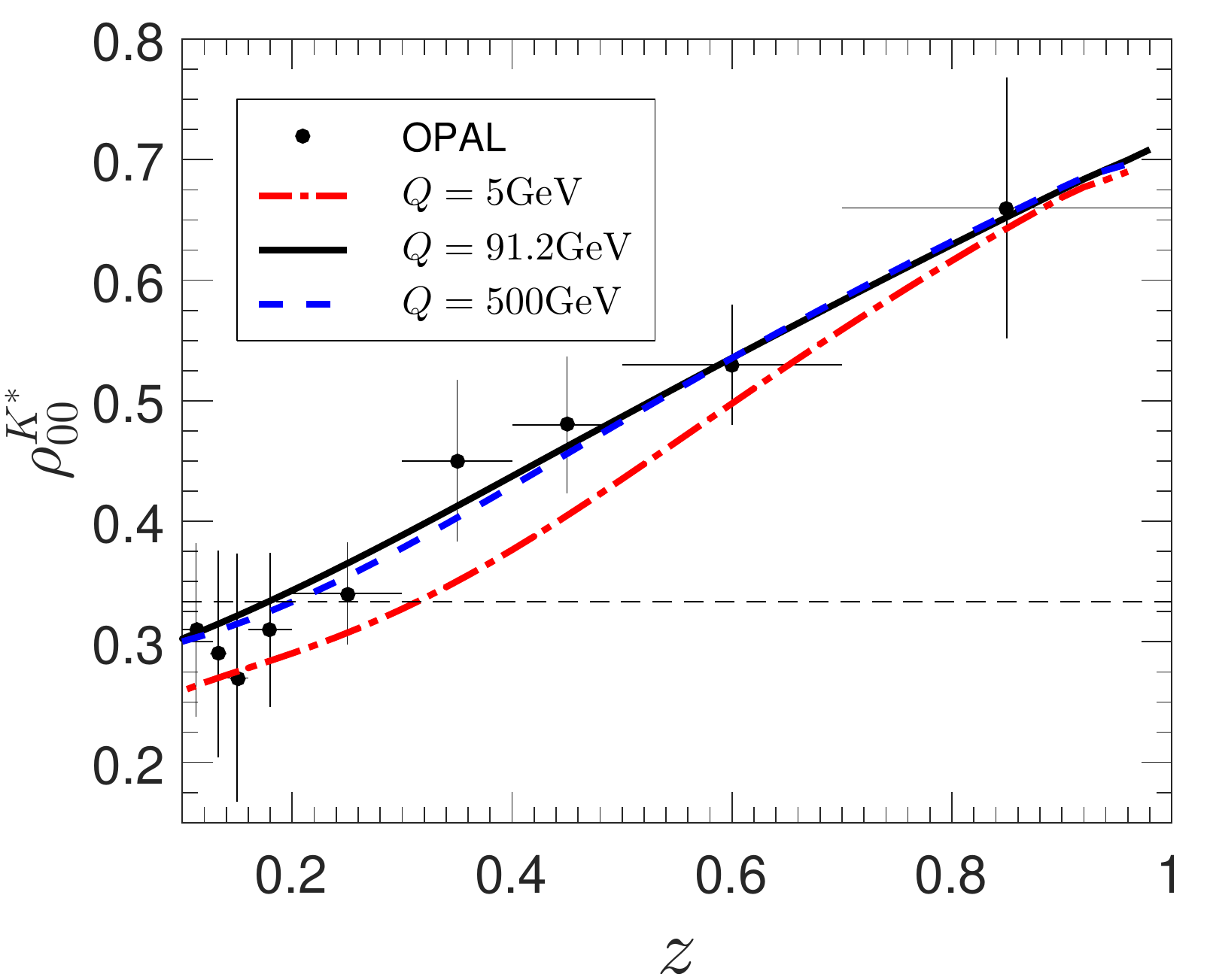}}
  \caption{(color on line). Results of energy dependence for (a) Lambda Hyperon longitudinal polarization, (b) $K^*$ spin alignment.} 
  \label{fig:pol} 
\end{figure}

At leading twist, the Lambda Hyperon longitudinal polarization and $K^*$ meson spin alignment are given by\cite{Chen:2016iey}
\begin{align}
& P_{L\Lambda} = \frac{\bar P_q W_q G_{1L}^{q\to\Lambda}(z,Q^2)}{W_q D_{1}^{q\to\Lambda}(z,Q^2)}, \\
& \rho_{00}^{K^*} = \frac{1}{3} - \frac{W_q D_{1LL}^{q\to K^*}(z,Q^2)}{3W_q D_{1}^{q\to K^*}(z,Q^2)}.
\end{align}
We make a simple parameterization for the polarized FFs at $Z^0$ pole energy,
and calculate the energy dependence, in which we have also considered the QCD evolution of the FFs.
The results are shown in Fig.~\ref{fig:pol}.
We see from Fig.~\ref{subfig:lambdapol} that $P_{L\Lambda}$ depends strongly on energy,
it reaches maximum at $Z^0$ pole and drops significantly at other energies, and vanishes at low energy region.
However, from Fig.~\ref{subfig:rho00} we see that $\rho_{00}^{K^*}$ changes little with energy, 
and still very large at low energies such as BELLE or BES.
This prediction can be checked in future experiments, 
which may promote our understanding of the hadron polarizations and hadronization mechanism.

\section{Summary} \label{sec:Summary}
We give a complete decomposition of TMD FFs from quark-quark correlator for spin-1 hadron.
A general kinematic analysis for $e^+e^- \to V\pi X$ is given, 
and the hadron azimuthal asymmetries and polarizations are expressed using the structure functions.
We also give the parton model results for the structure functions, azimuthal asymmetries and polarizations up to twist-3.
Numerical results of energy dependences of Hyperon longitudinal polarization and vector meson spin alignment are presented,
and the results show significant differences between them.

This work was supported in part by the National Natural Science Foundation of China
(Nos. 11675092 and 11375104),  the Major State Basic Research Development Program in China (No. 2014CB845406) 
and the CAS Center for Excellence in Particle Physics (CCEPP).



\begin{thebibliography}{0} 

\bibitem{Chen:2015tca} 
  K.~b.~Chen, S.~y.~Wei and Z.~t.~Liang,
  Front.\ Phys.\ (Beijing) {\bf 10}, no. 6, 101204 (2015)
  doi:10.1007/s11467-015-0477-x
  [arXiv:1506.07302 [hep-ph]].


    
\bibitem{Chen:2016moq} 
  K.~b.~Chen, W.~h.~Yang, S.~y.~Wei and Z.~t.~Liang,
  Phys.\ Rev.\ D {\bf 94}, no. 3, 034003 (2016)
  doi:10.1103/PhysRevD.94.034003



\bibitem{Goeke:2005hb} 
  K.~Goeke, A.~Metz and M.~Schlegel,
  Phys.\ Lett.\ B {\bf 618}, 90 (2005).


\bibitem{Wei:2013csa} 
  S.~y.~Wei, Y.~k.~Song and Z.~t.~Liang,
  Phys.\ Rev.\ D {\bf 89}, no. 1, 014024 (2014)
  doi:10.1103/PhysRevD.89.014024


 
\bibitem{Wei:2014pma} 
  S.~Y.~Wei, K.~b.~Chen, Y.~k.~Song and Z.~t.~Liang,
  Phys.\ Rev.\ D {\bf 91}, no. 3, 034015 (2015)



\bibitem{Pitonyak:2013dsu} 
  D.~Pitonyak, M.~Schlegel and A.~Metz,
  Phys.\ Rev.\ D {\bf 89}, no. 5, 054032 (2014)
  doi:10.1103/PhysRevD.89.054032


\bibitem{Gottfried:1964nx} 
  K.~Gottfried and J.~D.~Jackson,
  Nuovo Cim.\  {\bf 33}, 309 (1964).
  doi:10.1007/BF02750195



\bibitem{Boer:1997mf} 
  D.~Boer, R.~Jakob and P.~J.~Mulders,
  Nucl.\ Phys.\ B {\bf 504}, 345 (1997)
  doi:10.1016/S0550-3213(97)00456-2


\bibitem{Collins:1992kk} 
  J.~C.~Collins,
  Nucl.\ Phys.\ B {\bf 396}, 161 (1993)



\bibitem{Cahn:1978se} 
  R.~N.~Cahn,
  Phys.\ Lett.\ B {\bf 78}, 269 (1978).
  doi:10.1016/0370-2693(78)90020-5.


\bibitem{LambdaPol} 
  D.~Buskulic {\it et al.}  [ALEPH Collaboration],
  Phys.\ Lett.\ B {\bf 374}, 319 (1996).
  K.~Ackerstaff {\it et al.}  [OPAL Collaboration],
  Eur.\ Phys.\ J.\ C {\bf 2}, 49 (1998)


\bibitem{Rho00} 
  K.~Ackerstaff {\it et al.}  [OPAL Collaboration],
  Phys.\ Lett.\ B {\bf 412}, 210 (1997)
  P.~Abreu {\it et al.}  [DELPHI Collaboration],
  Phys.\ Lett.\ B {\bf 406}, 271 (1997).


\bibitem{Chen:2016iey} 
  K.~b.~Chen, W.~h.~Yang, Y.~j.~Zhou and Z.~t.~Liang,
  arXiv:1609.07001 [hep-ph].

\end{thebibliography}
\end{document}